\documentclass[aps,prd,nofootinbib]{revtex4-1}

\usepackage{graphicx}
\usepackage{enumerate}
\usepackage{amsmath}
\usepackage{amssymb}
\usepackage{amsfonts}
\usepackage{xcolor}
\usepackage{hyperref}
\usepackage{caption}
\usepackage[normalem]{ulem}

\hypersetup{colorlinks=true,linkcolor=redLinks,citecolor=greenLinks,
urlcolor=redLinks,
pdfborder={0 0 1}}
\hypersetup{
    bookmarks=true,         
    unicode=false,          
    pdftoolbar=true,        
    pdfmenubar=true,        
    pdffitwindow=false,     
    pdfstartview={FitH},    
    pdftitle={My title},    
    pdfauthor={Author},     
    pdfsubject={Subject},   
    pdfcreator={Creator},   
    pdfproducer={Producer}, 
    pdfkeywords={keyword1, key2, key3}, 
    pdfnewwindow=true,      
    colorlinks=false,       
    linkcolor=red,          
    citecolor=green,        
    filecolor=magenta,      
    urlcolor=cyan           
}

\newcommand{\bi}{\begin{itemize}}
\newcommand{\ei}{\end{itemize}}

\begin{document}
\title{Novel approach for the study of coherent elastic neutrino-nucleus scattering} 
\author{A. Galindo-Uribarri$^1$,$^2$} 
\email{uribarri@ornl.gov}\affiliation{$^1$~Oak Ridge National Laboratory, Oak Ridge, TN
  37831, USA}
 \affiliation{$^2$~Department of
  Physics and Astronomy, University of Tennessee, Knoxville, TN 37996,
  USA} 
  
\author{O. G. Miranda$^3$}
\email{omr@fis.cinvestav.mx} 
\author{G. Sanchez Garcia$^3$}
\email{gsanchez@fis.cinvestav.mx} \affiliation{$^3$~Departamento de
  F\'isica, Centro de Investigaci\'on y de Estudios Avanzados del IPN,
  Apdo. Postal 14-740, 07000 Ciudad de M\'exico, M\'exico.}

\begin{abstract}\noindent
      We propose the use of isotopically highly enriched detectors for
      the precise study of coherent-elastic neutrino-nucleus
      scattering (CEvNS).  CEvNS has been measured for the first time
      in CsI and recently confirmed with a liquid argon detector. It
      is expected that several new experimental setups will measure
      this process with increasing accuracy. Taking Ge detectors as a
      working example, we demonstrate that a combination of different
      isotopes is an excellent option to do precision neutrino physics
      with CEvNS, test Standard Model predictions, and probe new
      physics scenarios.  Experiments based on this new idea can make
      simultaneous differential CEvNS measurements with detectors of
      different isotopic composition. Particular combination of
      observables could be used to cancel systematic errors.  While
      many applications are possible, we illustrate the idea with
      three examples: testing the dominant quadratic dependence on the
      number of neutrons, $N$, that is predicted by the theoretical
      models; constraining the average neutron root mean square (rms)
      radius; and testing the weak mixing angle and the sensitivity to
      new physics.  In all three cases we find that the extra
      sensitivity provided by this method will potentially allow
      high-precision robust measurements with CEvNS and particularly,
      will resolve the characteristic degeneracies appearing in new
      physics scenarios.
    \end{abstract}

\maketitle

\section{Introduction}
Neutrinos in the energy region (1 MeV $\leq$ $E_{\nu}$ $\leq$ 50 MeV)
such as those generated in reactors or pion decay-at-rest sources
($\pi$-DAR), provide a new means of testing the Standard Model (SM)
and its possible extensions. Precise measurements are required to
understand the nature of the neutrino, to elucidate the phenomena that
give rise to its unique properties and to determine its impact on the
evolution of the Universe. After more than forty years of being
proposed~\cite{Freedman:1973yd}, the coherent elastic neutrino
nucleus-scattering (CEvNS) was observed for the first time in 2017 by
the COHERENT collaboration~\cite{Akimov:2017ade,Akimov:2018vzs}
through a CsI detector located at the Spallation Neutron Source (SNS)
at Oak Ridge National Laboratory (ORNL). This process has also been
confirmed by the same collaboration by using a liquid argon
detector~\cite{Akimov:2019rhz,Akimov:2020pdx}.

A precise measurement of CEvNS is of interest for many fields such as
nuclear physics, particle physics, and applications. Just to mention
some examples, CEvNS can help to extract detailed information about
the nuclear radius for different target
materials~\cite{Cadeddu:2017etk,Papoulias:2019lfi,AristizabalSierra:2019zmy,Coloma:2020nhf},
form factors~\cite{Hoferichter:2020osn}, precision
physics~\cite{Tomalak:2020zfh} as well as to constrain parameters
which describe physics beyond the SM such as non-standard
interactions~\cite{Barranco:2005yy,Scholberg:2005qs,Billard:2018jnl},
light mediators~\cite{Farzan:2018gtr}, neutrino magnetic
moments~\cite{Wong:2003xj,Miranda:2019wdy} or sterile
neutrinos~\cite{Dutta:2015nlo,Canas:2017umu,Miranda:2020syh}. An
application of considerable interest of reactor antineutrino detection
using CEvNS is in nuclear security~\cite{Bowen:2020unj}.
 
We propose a novel approach for the precise study of CEvNS; namely,
the use of isotopically highly enriched detectors.  We discuss Ge as a
concrete example to show feasibility, but the proposed technique of
using isotopically enriched material has much more general
applicability.  A detector array based on multi-isotope detectors
would have a particular combination of observables providing a fast
unambiguous signature of antineutrino detection.

It was noticed, since the first theoretical proposal of CEvNS by
Freedman~\cite{Freedman:1973yd}, that its corresponding cross section
has a quadratic dependence on the number of neutrons due to its
coherent character.  This enhancement makes the process very
attractive both experimentally and
theoretically~\cite{Barranco:2005yy}. The large cross section makes it
the dominant process at low energies and opens a new detection channel
that can give new independent physical information. Despite this
advantage, the heavy mass of the nucleus implies the detection of a
very low energy recoil and makes the measurement an experimental
challenge. Nuclear recoil signals at low energies are not well
calibrated and the ability to observe CEvNS from a reactor flux is
very dependent on low-energy quenching factors (QF). These are not
well known and their uncertainties are considerable. Significant
progress has been made recently to better understand the associated
systematic effects~\cite{Collar:2019ihs,Collar:2021fcl}.  An example
is the dramatic reduction of uncertainties on the QF's for CsI from
25\% in 2017 (first observation of CEvNS) to less than 5\% in
2021~\cite{Tayloe}.  Long experiments might require corrections due to
small drifts or for nonlinearities in the signal digitizer.  Besides,
as the process is testing a new energy region, it is natural that
theoretical uncertainties will appear. That is the case, for instance,
of the mean neutron radius of the target nuclei, for which, usually,
only theoretical predictions exist.  As a first step to test physics
beyond the SM, the variables under discussion have to be well
determined.

In this concept paper we propose a general experimental method to
improve the CEvNS accuracy by using an array of detectors of the same
element but different isotopes. The use of such an array will help to
diminish systematic errors (from flux uncertainties, for instance) and
to mitigate the dependence on form factors (FF), allowing a better
knowledge of the cross section and making it an even better tool for
testing new physics.  Currently, there are several proposals either
ongoing or planned to measure CEvNS with a stopped-pion neutrino
source or with reactor antineutrinos using different target materials.
For example, CONNIE using Si-based
CCDs~\cite{Aguilar-Arevalo:2016qen,Aguilar-Arevalo:2019jlr};
COHERENT~\cite{Akimov:2018ghi}, CONUS~\cite{Lindner:2016wff},
$\nu$GEN~\cite{Belov_2015}, and TEXONO~\cite{Wong:2005vg}, using
ionization-based Ge semiconductors; and MINER~\cite{Agnolet:2016zir},
NuCLEUS~\cite{Angloher:2019flc}, and RICOCHET~\cite{Billard:2016giu}
using cryogenic detectors (Si, Zn, Ge). The proposed technique
discussed here could be applied to any of these technologies, but to
illustrate the idea, we will only consider the germanium case.  The
process of isotopically enriching germanium is a well-developed
technology that has provided highly-enriched $^{76}$Ge for detectors
used in the search for neutrinoless double beta
decay~\cite{Gunther:1997ai,Gradwohl:2020kms}.  Furthermore,
isotopically modified Ge detectors depleted in $^{76}$Ge have been
produced and showed identical performance as those produced previously
from natural germanium
material~\cite{Budjas:2013bea,Agostini:2012ifa}.  Mitigation or
cancelation of the effect of systematic uncertainties is a crucial
issue in many fields of Physics. The high isotopic purity of 90 $\%$
or better that can be achieved in the detector materials (e.g. Ge) and
the possibility of producing identical detectors that differ only in
the number of neutrons provides a unique laboratory for the study of
the CEvNS interaction.

We present how such an array of Ge detectors could test and constrain
different parameters involving nuclear and particle physics.  In
section \ref{sec:N}, we study the $N^2$ CEvNS dependence. We study the
sensitivity in determining of the nuclear rms radius on section
\ref{sec:RMS}, which is relevant for the characterization of DM
detectors.  We discuss in section \ref{sec:SM-NSI} the expected
constraints for the weak mixing angle at low energies and Non-Standard
Interactions (NSI) parameters. In all the cases, we show the impact of
the correlation between systematic uncertainties on the determination
of these observables.  We focus our discussion on these examples, but
it is important to remark that there is a wide margin to drastically
improve other observables with this experimental approach.

Furthermore, it will be shown that the differential nuclear recoil
spectra have a non-trivial relative shape between the different
isotopes. Its understanding will also help to inform the $N$
dependence of CEvNS and other observables, by conveniently choosing
the appropriate region of energy.  We will show that this setup of
different germanium detectors can probe this rule.  As with any CEvNS
experiment, these measurements are inevitably affected by systematic
uncertainties that mainly result from quenching and form
factors. However, in this particular case, the system of coupled
detectors shares the same systematic effects and the correlations help
to make a cleaner statement about the relative value of the cross
sections.

Within the SM, the explicit form of the CEvNS cross section is 
\begin{equation} 
\left(\frac{d\sigma}{dT}\right)_{\rm SM}^{\rm coh} = \frac{G_{F}^{2}M}{\pi}\left[1-\frac{MT}{2E_{\nu}^{2}}\right]
 [Zg^{p}_{V}F_Z(q^2)+Ng^{n}_{V}F_N(q^2)]^{2} \label{eq:00}
\end{equation}
with $M$ the mass of the nucleus, $E_{\nu}$ the incident neutrino
energy, and $F_{X}$ the nuclear form factors, with $X = Z, N$ for
protons and neutrons, respectively.  The factors
$g_{V}^{p}=1/2-2\sin^2\theta_W$ and $g_{V}^{n}=1/2$ are the weak
coupling constants. Notice that $g_{V}^{p} << g_{V}^{n}$ and, in
consequence, the main dependence in this formula goes as $N^2$.

A precise confirmation, for instance, of the $N^2$ dependence is
challenging.  The uncertainties coming from very different detectors
with different exposure to the flux makes this task non-trivial. As we
show, the use of multiple enriched detectors under simultaneous
exposure time, would solve the issue of dealing with different
systematic uncertainties.  Indeed, for elements with multiple
isotopes, the $N^2$ dependence of CEvNS can provide a significant
variation in the measured number of events for each nuclei. In
particular, for germanium ($Z =$ 32) we have five stable natural
isotopes: $^{70}$Ge, $^{72}$Ge, $^{73}$Ge, $^{74}$Ge, and $^{76}$Ge,
implying a difference of up to $15$~\% in the relative number of
neutrons.
\begin{figure}[t] 
\begin{center}
\includegraphics[width=0.43\textwidth]{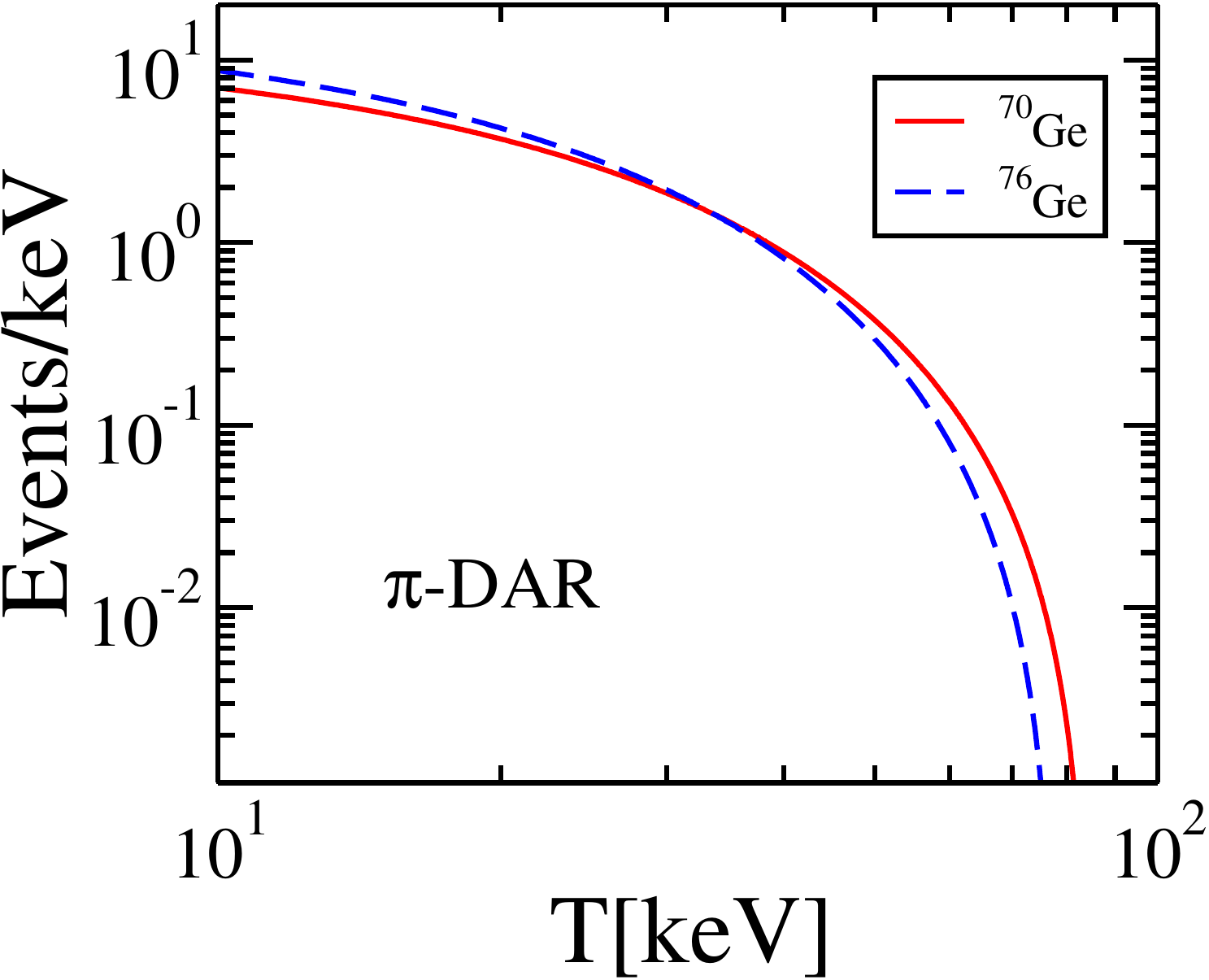}
\includegraphics[width=0.41\textwidth]{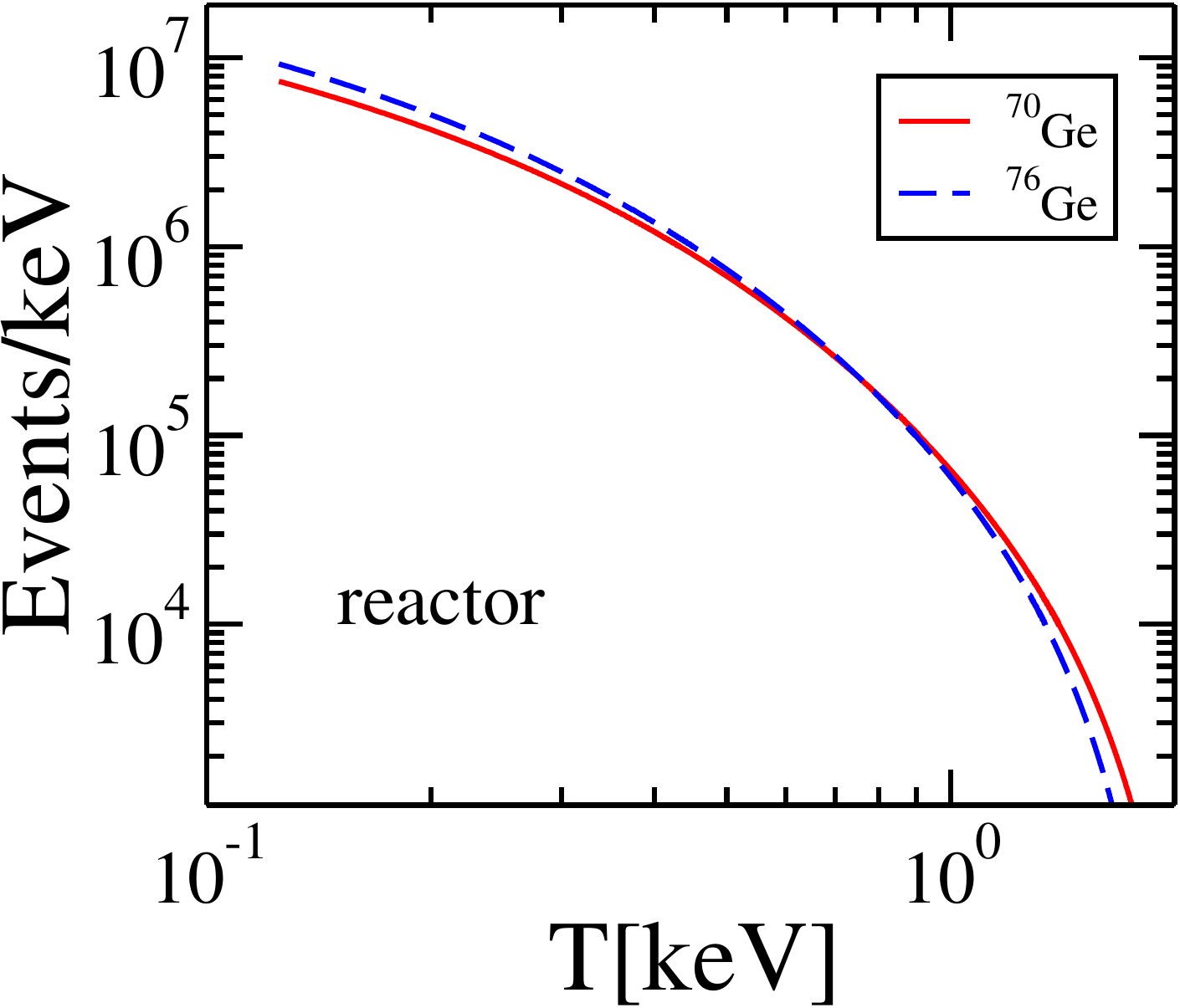}
\end{center}
  \caption{
  Differential event rate in terms of the nuclear recoil for two
  different isotopes ($^{70}$Ge and $^{76}$Ge) exposed to a stopped-pion neutrino source (left) or a reactor flux (right). As discussed in the
  text, we can see that for higher energies the lighter $^{70}$Ge has higher predicted rates per keV in both cases.} 
  
\label{fig:diffXsec}
\end{figure}
It is important to notice the behavior of the nuclear recoil energy
spectra due to the neutron number. We would expect that the main
difference between any two isotopes should be dominated by the
variation between the squares of the corresponding number of neutrons
for each isotope. However, the different masses of the isotopes also
play an important role, especially at the high energy tail of the
recoil energy spectrum. We can illustrate this by considering two
different isotopes, having $N_1$ and $N_2$ neutrons and a mass given
approximately by $(Z+N_i)m_N$ with $m_N$ the mass of an average
nucleon. Taking only leading terms, the difference between their cross
sections will be, approximately,

\begin{equation}
\delta = \frac{d\sigma(N_2)}{dT} - \frac{d\sigma(N_1)}{dT} =\frac{G_F m_N}{\pi}\left( a(N_1,N_2,Z) - b(N_1,N_2,Z) \frac{m_NT}{2E^2_\nu} \right)
\end{equation}
with
\begin{eqnarray}
  a(N_1,N_2,Z) &=& {g^n_V}^2 \left( Z(N^2_2 - N^2_1) + (N^3_2 - N^3_1) \right) \nonumber \\
  b(N_1,N_2,Z) &=& {g^n_V}^2 \left( Z^2(N^2_2 - N^2_1) + 2Z(N^3_2 - N^3_1)
  + (N^4_2 - N^4_1)  \right) . 
  \end{eqnarray}
For the case of $^{70}$Ge and $^{76}$Ge, we will have $a/b \simeq
0.01$, which means that the difference, $\delta$, will vanish for
$T=0.02E^2_\nu/m_N$. For example for a reactor antineutrino of
$E_\nu$ = {6~MeV} this happens around $T\approx0.7$~keV, while for a $\pi$-DAR
source for a neutrino  of $E_\nu$ = {40~MeV}, the same situation
arises at $T\approx 32$~keV. The more appealing result is that
$\delta$ will be negative for recoil energies above these values, that
is, despite our expectation of a higher number of events for heavier
isotopes, this is not the case in the tail of the recoil spectrum, an
odd feature that could help to test the predicted CEvNS cross
section. This is illustrated in Fig.~\ref{fig:diffXsec} for reactor
and $\pi$-DAR neutrinos, where we show the expected differential rate
for two different isotopes after integrating over the incoming
neutrino energy in each case.

In general, to study the response of the proposed array of
isotopically enriched detectors, we start by defining the expected
number of events, given by the convolution of the differential cross
section, $d\sigma/dT$, and the neutrino flux, $\phi(E)$, that is:
\begin{equation}
\mathcal{N}^{theo} = N_{D}\int_{T}A(T)dT\int_{E_{min}}^{E_{Max}}dE \phi(E) \frac{\mathrm{d} \sigma }{\mathrm{d} T},
\label{SNS_Events}
\end{equation}
\newline\newline with $N_{D}$ the effective number of target nuclei
within the detector during the running time of the experiment, and
$A(T)$ an acceptance function.

We can use the total number of events predicted by
Eq.~(\ref{SNS_Events}) to perform different tests. In general,  given a 
set of different detectors for which correlations are present, we
can use a simple $\chi^{2}$ analysis by minimizing the function:   

\begin{equation}
\chi^2 = \sum_{ij}(\mathcal{N}_i^{theo} - \mathcal{N}_i^{exp}) [\sigma_{ij}^{2}]^{-1}
(\mathcal{N}_j^{theo} - \mathcal{N}_j^{exp}) ,
\label{chi2}
\end{equation}
where $\sigma_{ij}^{2}$ is the covariant matrix and
$\mathcal{N}_{i}^{exp}$ the `experimental' measurement, which we will
take as the SM prediction in each case. The theoretical number of
events, $\mathcal{N}^{th}_{i}$, depends on the set of parameters under
study, and the indices $i$, $j$ run over the number of detectors
within the array.

To illustrate the idea, we can consider an experimental array of three
different germanium isotopes: $^{\ell}$Ge, $^{m}$Ge, and $^{n}$Ge
(with $\ell$, $m$, and $n$ the total number of nucleons), located at
the same distance from the neutrino source and taking data
simultaneously.  Under these conditions, the detectors will be exposed
to the same neutrino flux, regardless of any variations through the
running time, and there will be correlations among the systematic
uncertainties which, as we will see, will be useful to test the
observable under study.  This approach can easily be extended to more
isotopes and different materials. In our particular example, the
covariance matrix is a $3\times 3$ matrix with nonzero elements
outside the diagonal due to the correlation between the detectors.
Denoting by $A$ and $B$ the most significant sources of errors within
the experiment, with $\sigma^A_{k}$ and $\sigma^B_{k}$ ($k = \ell, m,
n$) their corresponding uncertainties for each isotope, we have:
\begin{equation}
\sigma^{2} = 
\begin{pmatrix}
{\sigma_{\ell}^{stat}}^2 + {\sigma _{\ell}^{A}}^2 + {\sigma_{\ell}^{B}}^2 & \sigma_{\ell}^{A}\sigma _{m}^{A} +  \sigma_{\ell}^{B}\sigma_{m}^{B} &\sigma_{\ell}^{A}\sigma_{n}^{A} +  \sigma_{\ell}^{B}\sigma_{n}^{B}\\ 
\sigma_{\ell}^{A}\sigma _{m}^{A} +  \sigma _{\ell}^{B}\sigma _{m}^{B} & {\sigma _{m}^{stat}}^2 + {\sigma _{m}^{A}}^2 + {\sigma _{m}^{B}}^2 & \sigma _{m}^{A}\sigma _{n}^{A} +  \sigma _{m}^{B}\sigma _{n}^{B}\\ 
\sigma_{\ell}^{A}\sigma _{n}^{A} +  \sigma _{\ell}^{B}\sigma _{n}^{B}&\sigma _{m}^{A}\sigma _{n}^{A} +  \sigma _{m}^{B}\sigma _{n}^{B}  & {\sigma _{n}^{stat}}^2 + {\sigma _{n}^{A}}^2 + {\sigma _{n}^{B}}^2 
\end{pmatrix} .
\label{c-matrix}
\end{equation}
\newline\newline This notation allows us to generalize the matrix
elements for experiments with different neutrino sources and detection
technologies.  The superscript in each case denotes the systematic
uncertainty source, while the sub-index makes reference to the
associated isotope.  In what follows, we will show how this particular
approach can be used to study different parameters regarding nuclear
and particle physics.

\section{Testing the $N^{2}$ dependence of CEvNS}\label{sec:N} 

The dominance of the CEvNS cross-section, when compared to other
processes, comes from the characteristic $N^2$ dependence. Here we
show some predictions consistent with this dependence when comparing
the relative measurements between detectors that differ only in the
number of neutrons. To do so, we replace the $N$ factor in
Eq.~(\ref{eq:00}) by a variable, $N'$, which will test how much an
experimental measurement deviates from the $N^{2}$ dependence. Just as
the number of neutrons, this factor will be different for each
nucleus, and we can express the predicted number of events for a
particular detector as:

\begin{figure}[t] 
\begin{center}
\includegraphics[width=0.432\textwidth]{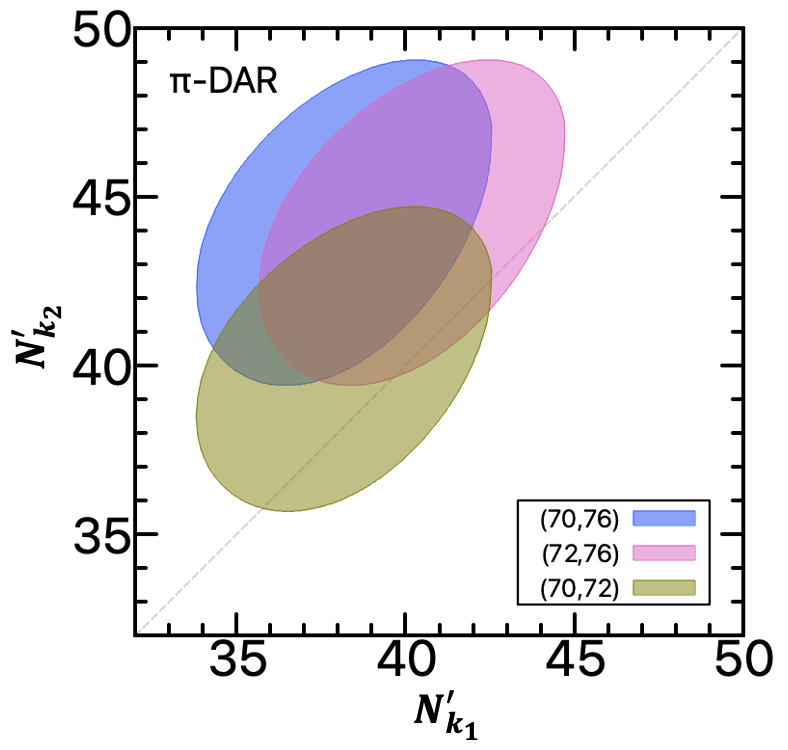}
\includegraphics[width=0.43\textwidth]{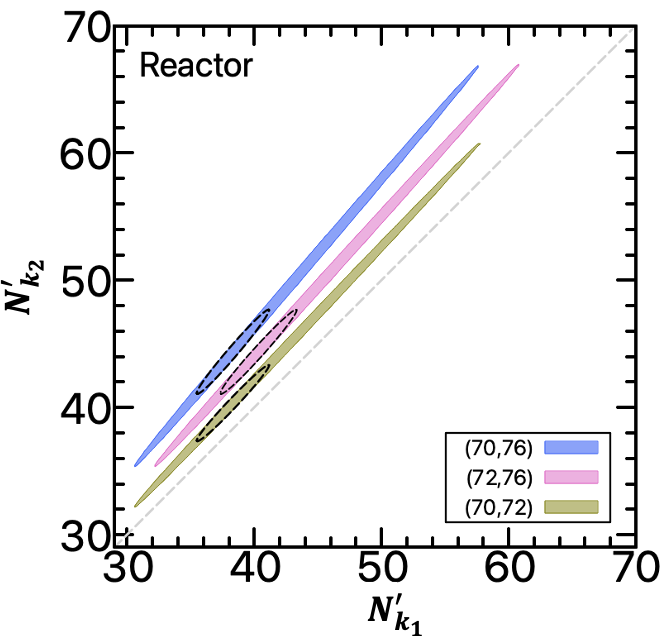}
\end{center}
\caption{Left panel: Expected measurement of the $N$ dependence in 
  the case of
  different Germanium detectors within the same experimental setup. We
  assume that the three detectors, with equal mass, will be exposed to
  the same $\pi$-DAR neutrino beam with equal contribution of $\sigma_{k}^{A} =$ $\sigma_{k}^{B} =$ 5\% from the main systematic errors (see the text
    for details), that the systematic errors will be correlated, and
  we consider events from $5$ to $30$~keV nuclear recoil.  We show the allowed regions for $N'$ values by pairs of isotopes with total nucleons ($k_{1}, k_{2}$).The green, magenta and blue
  regions correspond, respectively, to the results of $^{70}$Ge {\it vs} $^{72}$Ge, $^{72}$Ge {\it vs} $^{76}$Ge and $^{70}$Ge {\it vs} $^{76}$Ge at a $90$~\%
  C.L. when marginalizing over the missing detector. The first number in the ordered pair on the inset corresponds to the horizontal axis and the second to the vertical axis.  Right panel: We show for comparison the case of a reactor based experiment with an electron recoil energy from 1 to 2 keV.  Here we explicitly show the comparison of the results for different systematic uncertainties assumptions.  Colored regions refer to the pessimistic case of $\sigma_{k}^{A} =$ 25\% and $\sigma_{k}^{B} =$ 10\% main systematic contributions. Dashed contours refer to the case of 5\% equal contributions.}
  \label{fig:Ge1}
\end{figure}

\begin{equation}
\begin{aligned}
\mathcal{N}^{theo} &= N'^{2}\left ( g^{n^{2}}_{V}N_{D} \frac{G_{F}^{2}M}{\pi}\int_{T}A(T)dT\int_{E_{min}}^{E_{Max}}dE\left[1-\frac{MT}{2E_{\nu}^{2}}\right] \phi(E) F_N^{2}(q^2)\right )  \\
& +  N'\left (Zg^{p}_{V}g^{n}_{V}N_{D}\frac{G_{F}^{2}M}{\pi}\int_{T}A(T)dT\int_{E_{min}}^{E_{Max}}dE\left[1-\frac{MT}{2E_{\nu}^{2}}\right] \phi(E)F_Z(q^2)F_N(q^2)\right )\\
& + Z^{2}g^{p^{2}}_{V}N_{D}\frac{G_{F}^{2}M}{\pi}\int_{T}A(T)dT\int_{E_{min}}^{E_{Max}}dE\left[1-\frac{MT}{2E_{\nu}^{2}}\right] \phi(E) F_Z^{2}(q^2)
\label{Approx3}
\end{aligned}
\end{equation}
\newline\newline The first term in Eq.~(\ref{Approx3}) explicitly
shows the dominant quadratic dependence on the number of neutrons for
$N' = N$. An $N^{2}$ rule would be absolute if $g^{p}_{V} <<
g^{n}_{V}$ so that the second and third terms would be
negligible. Indeed, this is close to the real case since $g^{p}_{V} /
g^{n}_{V} \approx 0.05$; nevertheless, in all our computations we
consider the complete expression of Eq.~(\ref{Approx3}).  We have
explicitly checked that our results for the limiting case $g^{p}_{V}
\to 0$ are qualitatively similar.  To test the sensitivity to the
$N^2$ rule we can perform a $\chi^{2}$ analysis following the
covariant matrix approach introduced in the previous section, where
the parameters under study will be the $N'_{k}$ factors, where the sub
index $k$ has been introduced to remark that $N'$ is different for
each isotope.  In all our computations we will assume an array of the
three different isotopes: $^{70}$Ge, $^{72}$Ge, and $^{76}$Ge, so we
take $\ell = 70$, $m = 72$, and $n = 76$ in
Eq. (\ref{c-matrix}). Other combinations can be taken among all of the
stable isotopes of germanium.  Regarding systematic uncertainties, for
$\pi$-DAR neutrinos we take the most significant contributions for
systematic uncertainties as those coming from quenching factors ($A$ =
QF) and form factors ($B$ = FF). Furthermore we consider two
scenarios, one where these contributions are of $\sigma_{k}^{A} =$
25\% and $\sigma_{k}^{B} =$ 10\%, respectively, and another scenario
where both contributions are of 5\%.  For reactor sources, we take the
main sources of systematic uncertainties as those coming from the
quenching factor ($A$ = QF), and the neutrino flux itself ($B$ = NF),
with the same numerical configurations as in the case of $\pi$-DAR
sources. In addition, for all our computations we consider background
effects by adding a 10\% contribution of the SM prediction to the
statistical uncertainty.

We consider now a neutrino flux from a stopped-pion neutrino source;
specifically one with similar characteristics to the SNS at ORNL
\cite{Akimov:2015nza}.  In such case, the total incident neutrino flux
is given by three different contributions which result from the decay
of pions and muons. To calculate the expected number of events, we
consider a mass of 10 kg for each isotope, a time of exposure of one
year, a Helm distribution for protons and neutrons, and an acceptance
function equal to a step function.

The $\chi^2$ function in Eq.~(\ref{chi2}) is a three variable function
that defines a volume in the parameter space $N'_{k}$ for the allowed
values of these factors which are consistent with the theory at a
desired C.L.. Three projections are shown in the left panel of
Fig.~(\ref{fig:Ge1}) at a 90\% C.L.  for the configuration where both
systematic uncertainties contribute with a 5\%.  For instance, the
magenta region shows the simultaneously allowed $N'_{k}$ values for
$^{72}$Ge and $^{76}$Ge when marginalizing the information about
$^{70}$Ge.  As we can see, the expected region forms an ellipse that
restricts the values of $N'_{k}$ to lie in correlation to one another
according to the $N^2$ rule. A similar analysis is shown in green for
the pair $^{70}$Ge, $^{72}$Ge and in blue for the pair $^{70}$Ge,
$^{76}$Ge.  Notice that the centroid of the regions are separated from
the diagonal, in proportion to the difference in neutron number, as
expected.

\begin{figure}[t] 
\begin{center}
\includegraphics[width=0.32\textwidth]{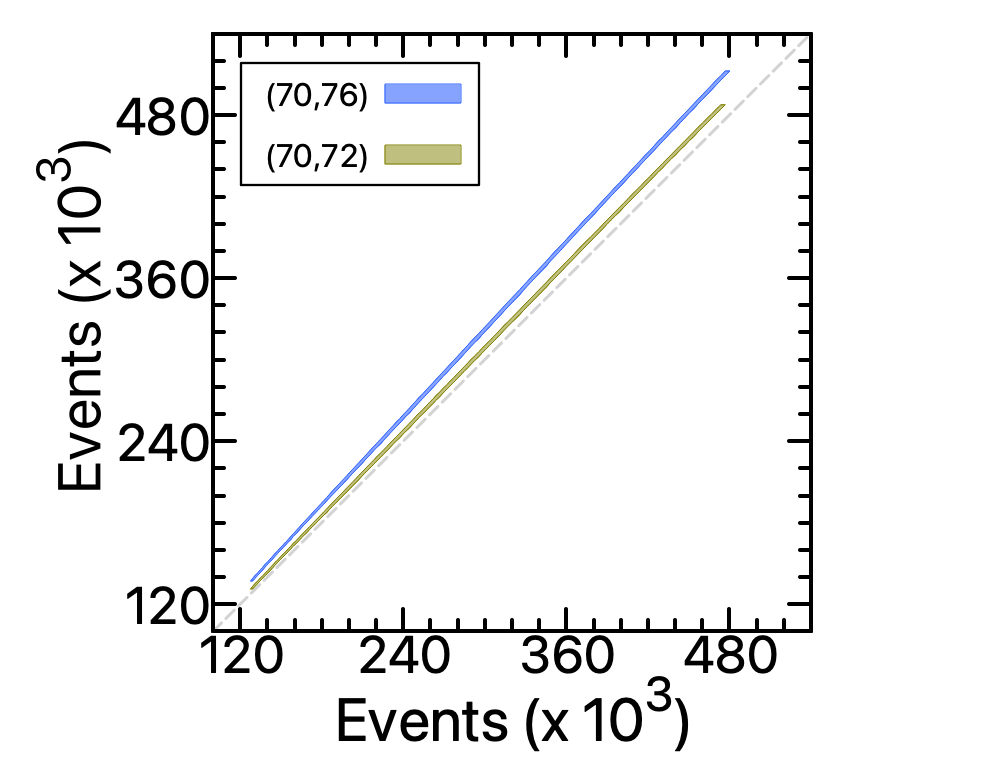}
\includegraphics[width=0.32\textwidth]{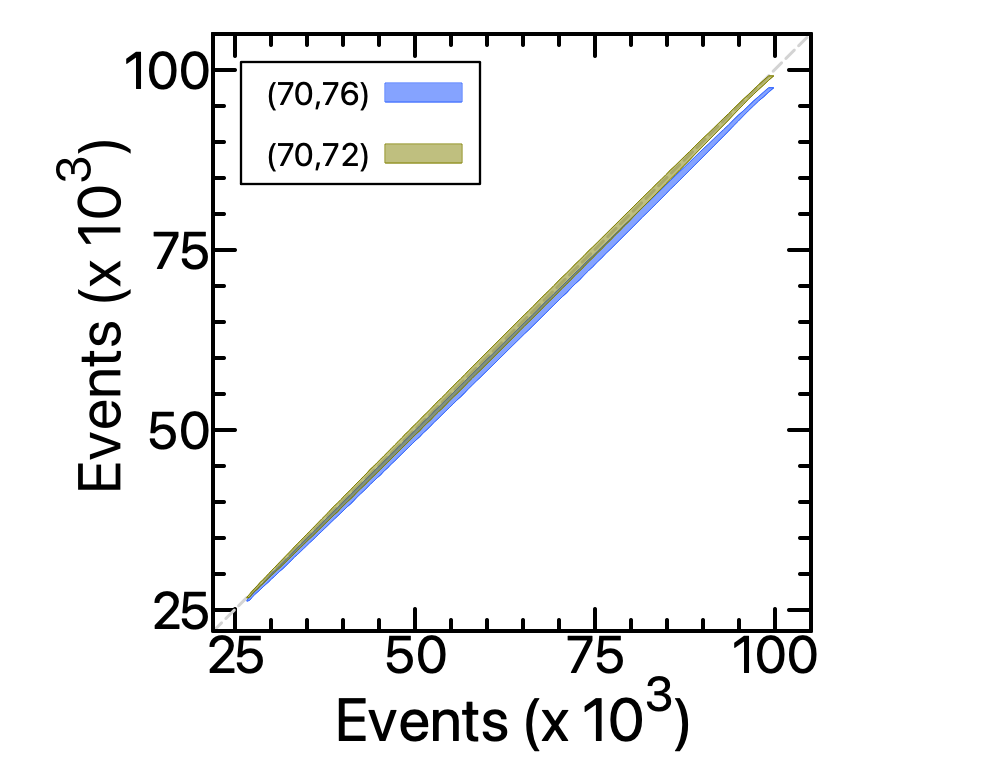}
\includegraphics[width=0.32\textwidth]{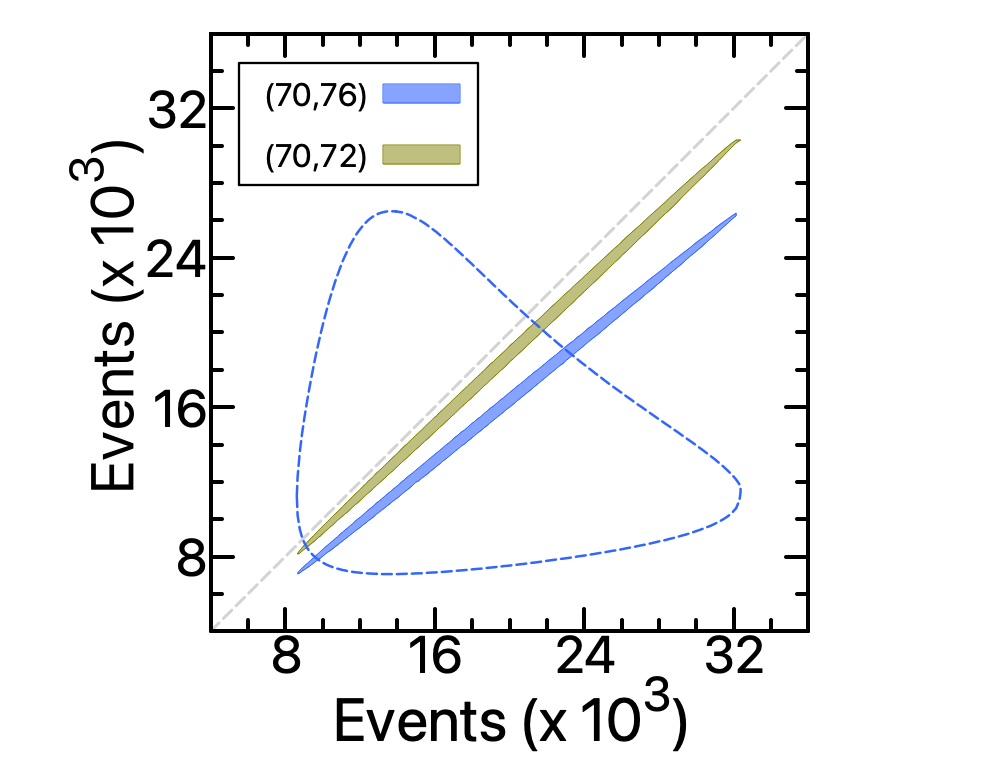}
\includegraphics[width=0.32\textwidth]{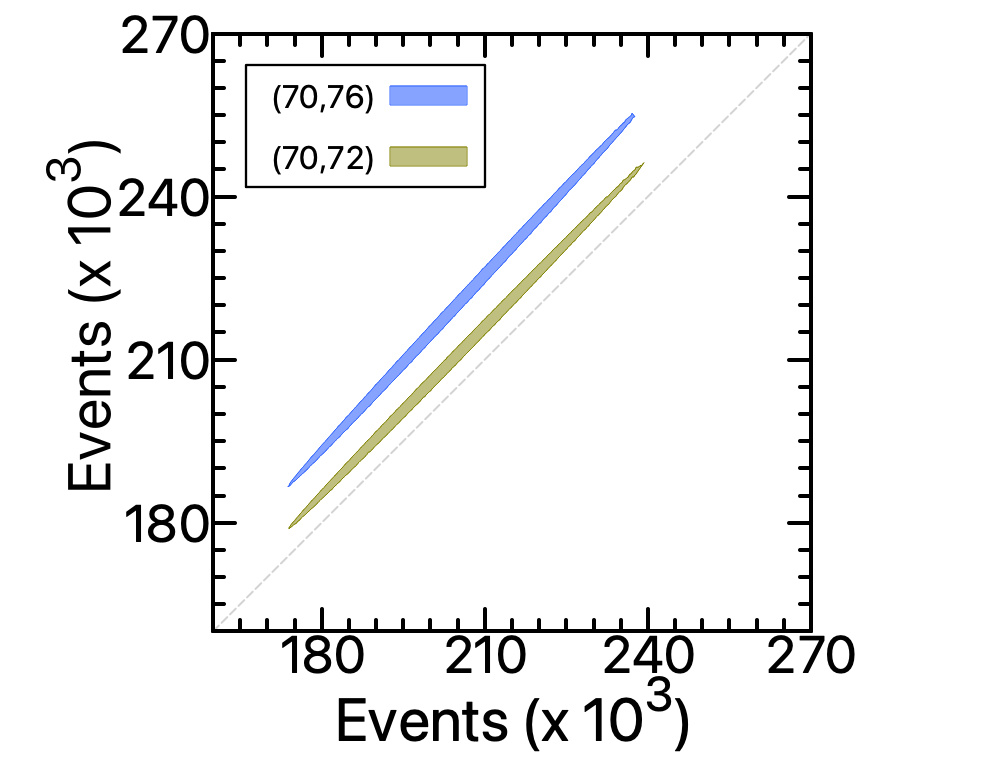}
\includegraphics[width=0.32\textwidth]{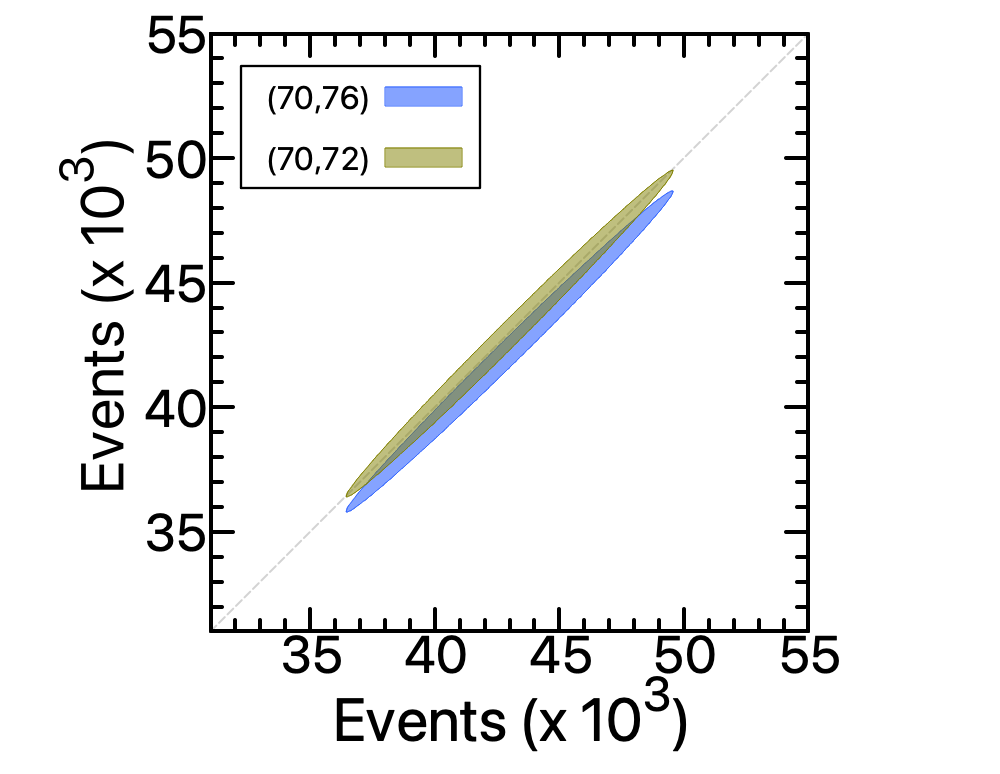}
\includegraphics[width=0.32\textwidth]{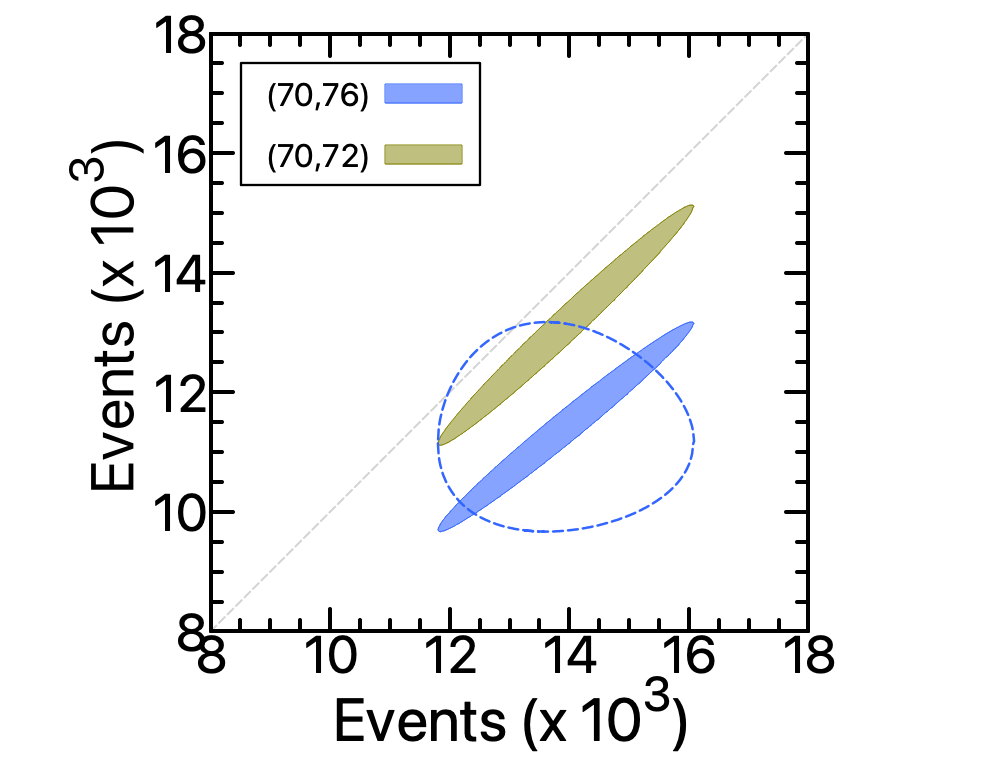}
\end{center}
\caption{Expected number of events for the three different Ge
  detectors exposed to a reactor antineutrino flux. The rate from the higher-mass of the two isotopes is shown on the vertical axis. Top panels refer to systematic uncertainties of $\sigma_{k}^{A} =$ 25\% and $\sigma_{k}^{B} =$ 10\%, while lower panels correspond to 5\% each. We show in the
  the left panels  the result of
  integrating on the low recoil energy of the spectrum from $0.4$ to
  $0.8$~keV.  In the central panels we consider an intermediate energy range from $0.7$ to
  $1$~keV, and the right panels a high energy range from $1$ to $2$~keV.  It can be seen that
 the the blue and green ellipses get 
  interchanged as expected from the energy spectrum shape shown in Fig.  \ref{fig:diffXsec}.  For comparison, the uncorrelated cases for the $^{70}$Ge {\it vs} $^{76}$Ge pair are also shown in the right panels (dashed contours). }
  \label{GeEvents}
\end{figure}

We also consider the case of the same detector array, exposed now to a
reactor antineutrino
flux~\cite{Huber:2004xh,Mention:2011rk,Kopeikin:1997ve}. In this case,
the flux from the reactor will be composed of only electron
antineutrinos. For our computations we assume a neutrino flux of
1$\times 10^{13}$ $\nu / cm^2 / s$ (a typical flux for several CEvNS
experiments~\cite{Lindner:2016wff}), one year of data taking and a
mass of 10 kg for each isotope. The incoming flux will be higher in
this case, although the nuclear recoil energy threshold is more
demanding. Despite this, if a $1$ keV recoil-energy threshold level
can be reached, it is expected that this set of detectors can test the
$N^2$ rule with good significance, as shown in the right panel of
Fig.~(\ref{fig:Ge1}), where we considered 1 keV $< T < $ 2 keV. We see
a clear improvement in the resolution for this rule. For illustration,
we show the two considered configurations of systematic uncertainties.
The colored regions represent the scenario for which the main
contributions to the systematic uncertainties have an impact of 25\%
and 10\%, while the dashed contours represent the corresponding case
where each of them contribute with a 5\%. We can see that the larger
systematic errors have an impact on the length of the ellipses, while
the width is still dominated by the statistical error.

We now comment on the viability of this array to test the predictions
of the CEvNS cross section regarding the number of events for a given
experiment. To this end we can refer to Fig. \ref{fig:diffXsec}, where
we see that the number of events in each detector depends on the
nuclear recoil energy region chosen.  Taking advantage of the high
statistics data that can be obtained with a reactor source, we can
explore the expected results for three different regions under the
same previously discussed conditions.  We show in
Fig.~(\ref{GeEvents}) the regions where we expect to have a
measurement of the number of events when comparing the data from
different detectors by pairs, again marginalizing the information of
the third detector, and considering the correlation effects for
different nuclear recoil energy intervals. Indeed, if the $N^{2}$ rule
holds, the different measurements in the germanium detectors will be
correlated. The top panels refer to the conservative case of large
systematic uncertainties, while the lower panels refer to the case
where each contributes with a 5\%.  If the detector technology is such
that allows a very low nuclear recoil energy threshold, then we get a
relation as seen in the left panels of Fig. (\ref{GeEvents}), where an
energy region between $0.4$ and $0.8$~keV has been assumed. We can
distinguish two clearly separated regions due to the high statistics
of events.  In the central panels we show the intermediate energy
region corresponding to $0.7-1.0$~keV, showing an overlap of the
predicted regions since we are considering an energy interval
corresponding to a neighborhood of the crossing point of the
differential spectrum shown in the right panel of
Fig. (\ref{fig:diffXsec}). Finally, in the right panels of
Fig.~(\ref{GeEvents}), we take a recoil energy spectrum from 1 keV to
2 keV, and the ordering of the ellipses is interchanged since the
lightest isotope detector will measure more events than the heavier
one. Even in this case the sensitivity would be enough to clearly
distinguish different pairs of measured event rates. For comparison,
the dashed contours show the result of the expected measured number of
events for the pair $^{70}$Ge, $^{76}$Ge in the case where we do not
have any correlations. It is clear that the effect of correlation is
of the highest importance.~\footnote{For this reactor example, the
  error in the antineutrino flux spectrum will also be a correlated
  error. However, it is expected that this error will be smaller than
  the dominant QF effects.} This impressive resolution could be also
useful to strongly constrain physics beyond the Standard Model,
especially if we can forecast a difference in the spectral shape of
the events. The above discussion applies for both configurations of
systematic uncertainties presented so far, the only difference being
the length of the axis of the ellipses when we consider the different
correlations.

\section{Constraining the neutron RMS radius.}
\label{sec:RMS}

\begin{figure}[t] 
\begin{center}
\includegraphics[width=0.31\textwidth]{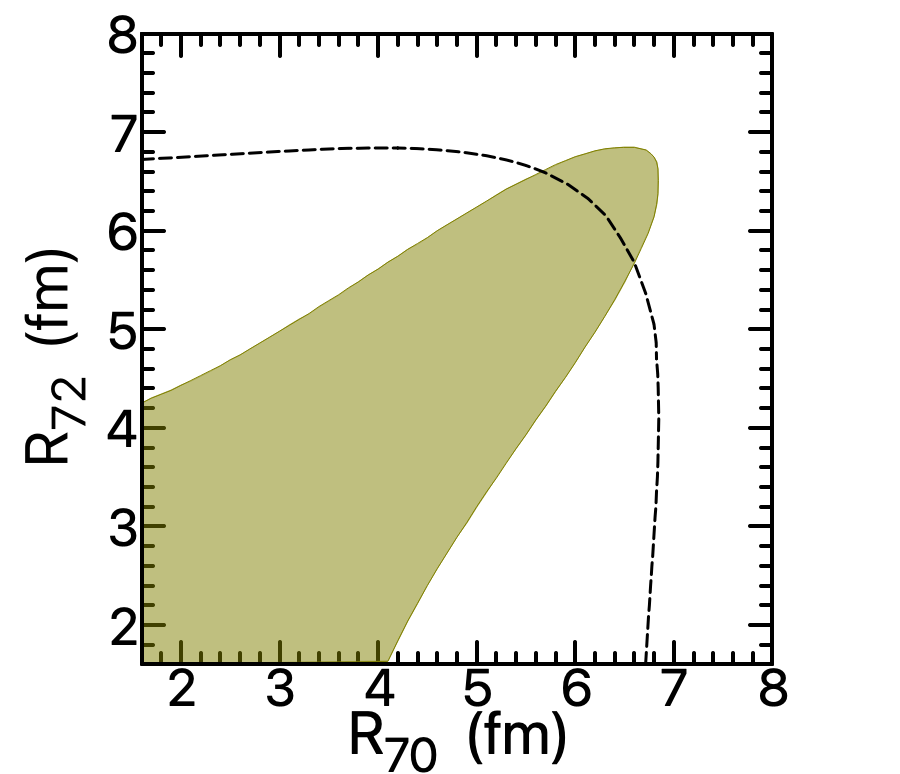}
\includegraphics[width=0.31\textwidth]{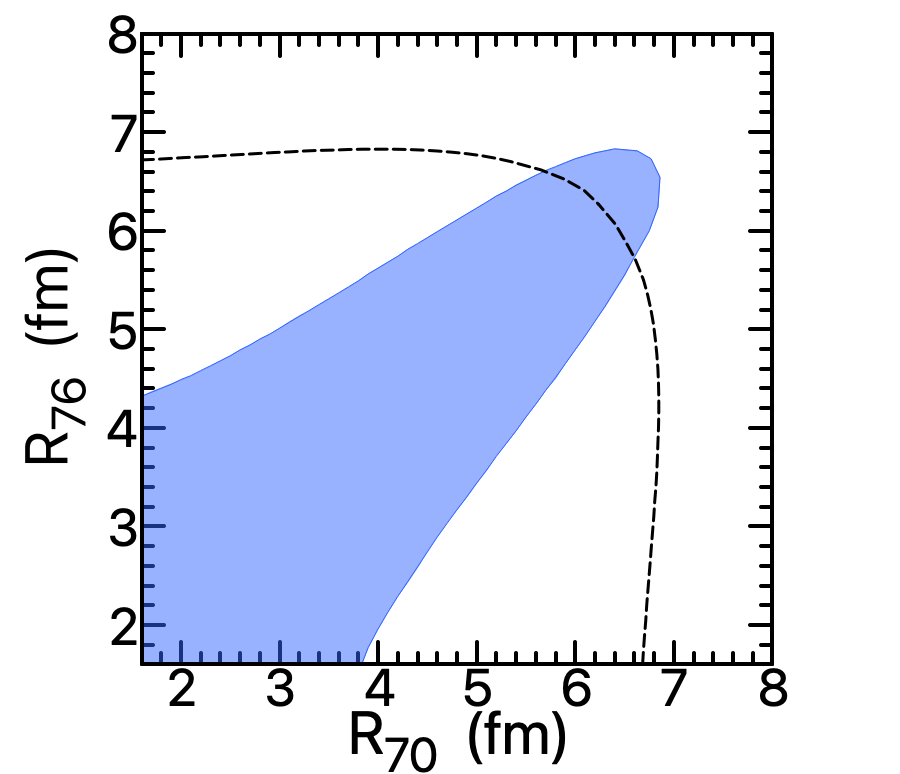}
\includegraphics[width=0.31\textwidth]{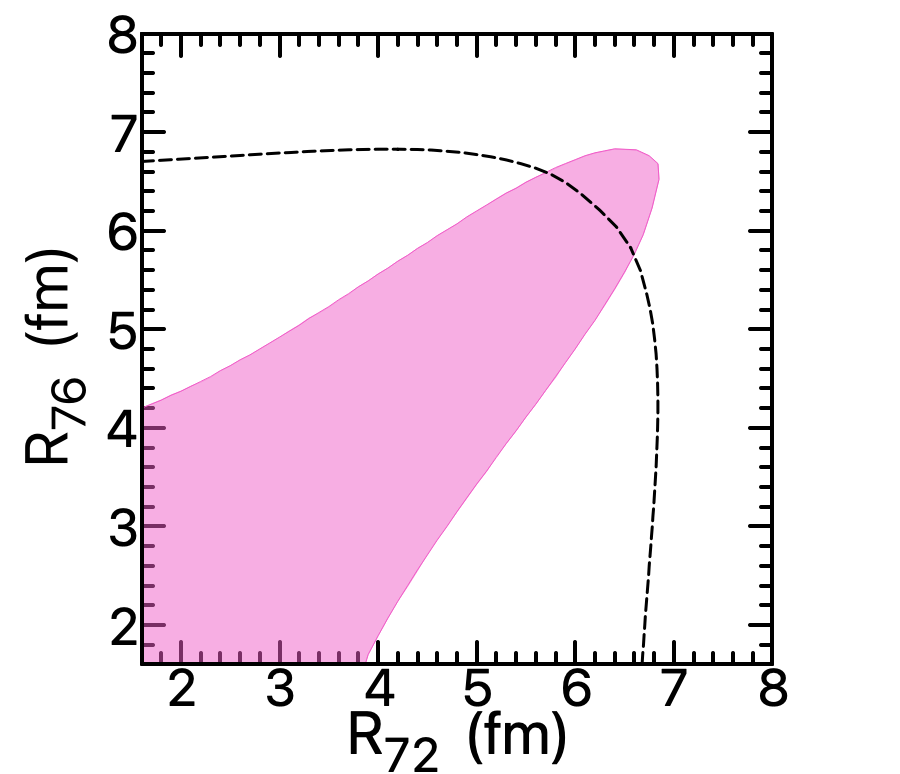}
\includegraphics[width=0.31\textwidth]{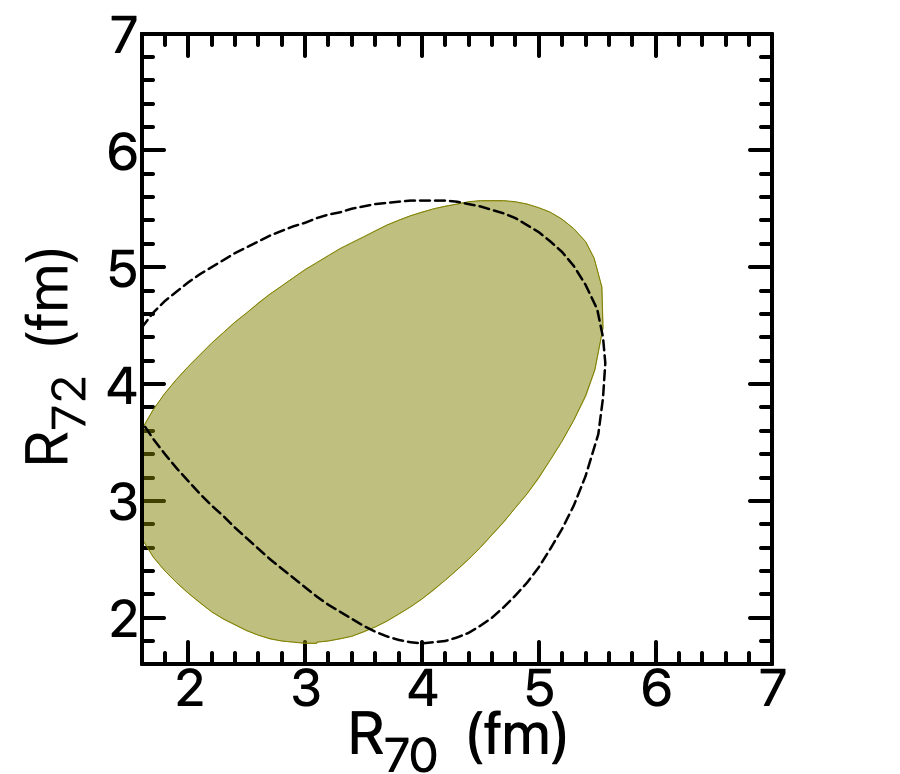}
\includegraphics[width=0.31\textwidth]{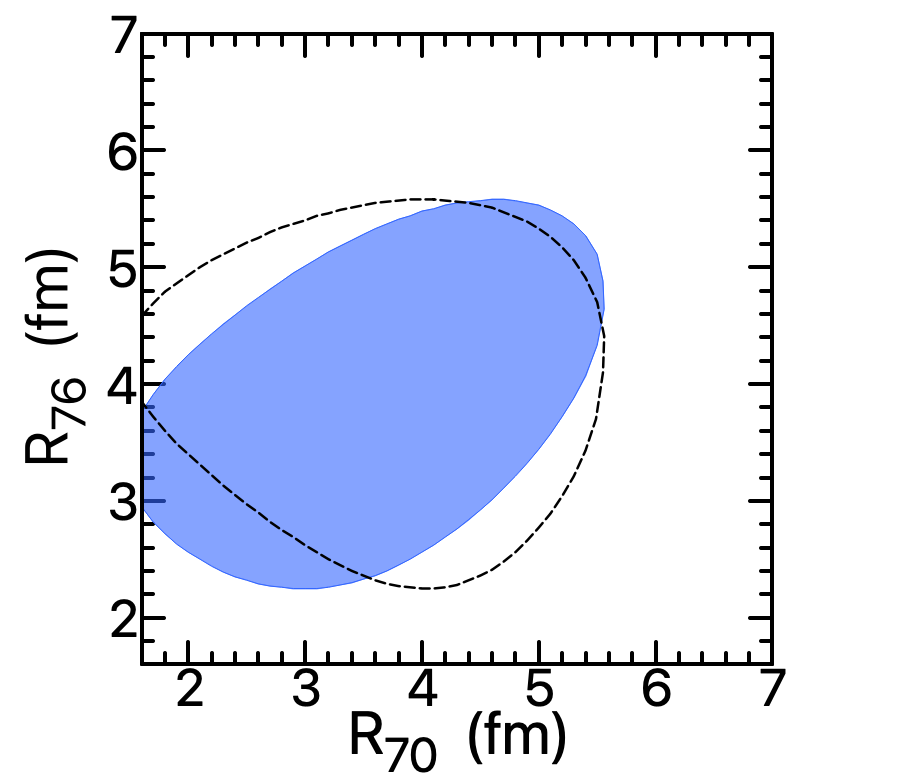}
\includegraphics[width=0.31\textwidth]{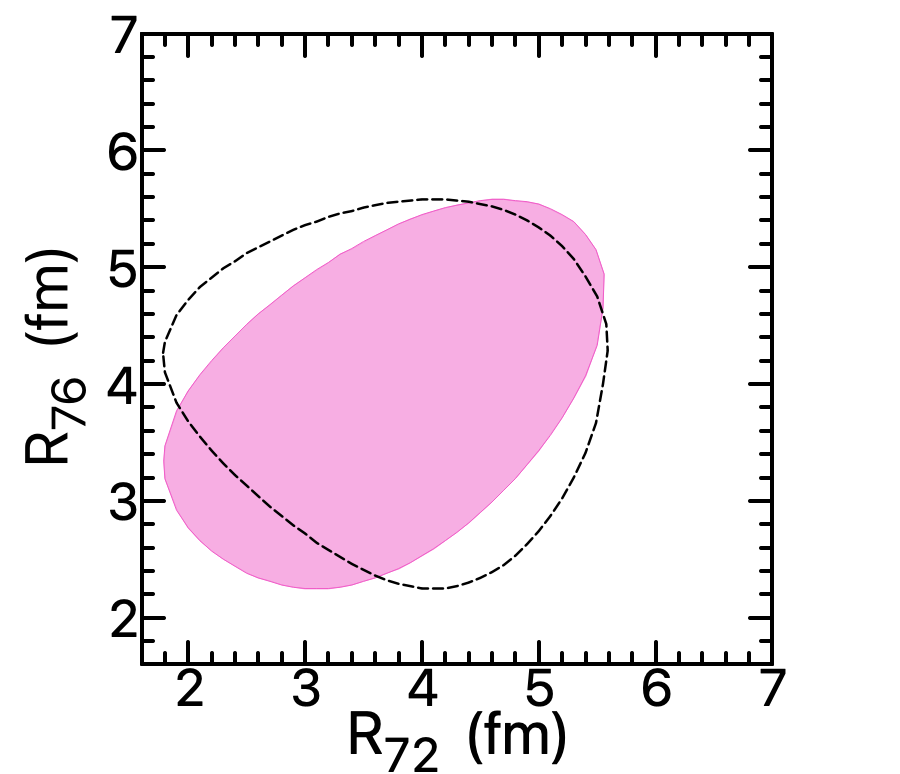}
\end{center}
\caption{ Neutron rms radius at a 90\% C.L.  for the pairs $R_{70}$ vs $R_{72}$ (left), $R_{72}$ vs $R_{76}$ (center), and $R_{70}$ vs $R_{76}$ (right), after marginalizing the information of the third isotope. In all cases the filled region represents the results when considering the correlation between detectors. The dashed line indicates the uncorrelated case. We consider background effects by adding a 10\% contribution of the SM prediction to the statistical uncertainty. Top panels refer to systematic uncertainties contributions of $\sigma_{k}^{A} =$ 25\% and $\sigma_{k}^{B} =$ 10\% , while lower panels correspond to 5\% each.
  \label{fig:RMS}}
\end{figure}

As we have pointed out, neutrinos coming from $\pi$-DAR sources are on
an energy regime where the nuclear form factors play an important role
in the theoretical prediction of the total number of events for a
given experiment. This dependence can be used in order to constrain
parameters relevant to nuclear physics as the neutron rms radius
($R_{n}$), defined as the average radius within which neutrons are
distributed in the nucleus. An accurate determination of this
parameter is of general interest not only for nuclear physics, but
also in the characterization of Dark Matter detectors and in the
experimental measurement of the neutrino floor. Currently, the neutron
RMS radius has only been experimentally measured for a few elements,
and the process of CEvNS can be used as a tool to determine the RMS
radius of the target material. Explicitly, in the Helm parametrization
of the form factor we have:

\begin{equation}
F_{N}^{Helm}(q^{2}) = 3\frac{j_{1}(qR_{0})}{qR_{0}}e^{\frac{-q^{2}s^{2}}{2}}
\label{HF}
\end{equation}
where $j_{1}(x)$ is the spherical Bessel function of order one,  and $R_{0}$ satisfies:

\begin{equation}
R_n^{2} = \frac{3}{5}R_{0}^{2} + 3s^{2}
\label{Rc2}
\end{equation} 
with $s$ the neutron skin. Our proposed experimental array can be used
to constrain $R_{n}$ for the involved isotope targets. To this end, we
use the same set of germanium detectors under the same conditions of
the analysis presented in section \ref{sec:N} for the SNS case.  We
perform the $\chi^{2}$ analysis, this time by considering the three
neutron rms radii as our variable of interest. Fig.  (\ref{fig:RMS})
shows the results at a 90\% C.L.  of the allowed regions for different
radii by pairs when marginalizing the information about the other
radius.  Top panels refer to large contributions from systematic
errors and lower panels to small contributions as in the previous
section. Left panel shows the allowed values of $R_{70}$ and $R_{72}$
after marginalizing the information of $R_{76}$.  The green region
shows the results when we consider the correlations between systematic
uncertainties, while the dashed line in the same figure represents the
situation where no correlations are considered.  The central panel of
the same figure shows the allowed region for the pair $R_{70},
R_{76}$, and the right panel shows the combination
$R_{72},R_{76}$. Here we have considered the background effects as the
10\% of the SM prediction in the number of events.  Regardless of the
configuration of systematic uncertainties, the effects of the
correlations are present to a certain degree.  Notice that,
physically, $R_{n} > 1.56$ fm according to Eq. (\ref{Rc2}).  Different
parameterizations exist for the distribution of both protons and
neutrons within the nucleus. Another common one in the literature is
the Symmetrized Fermi (SF) distribution. We have explicitly checked
that similar results can be found with this case.

\section{Constraining the Standard Model and non-standard interactions parameters} \label{sec:SM-NSI}

 \begin{figure}[t] 
\begin{center}
\includegraphics[width=0.8\textwidth]{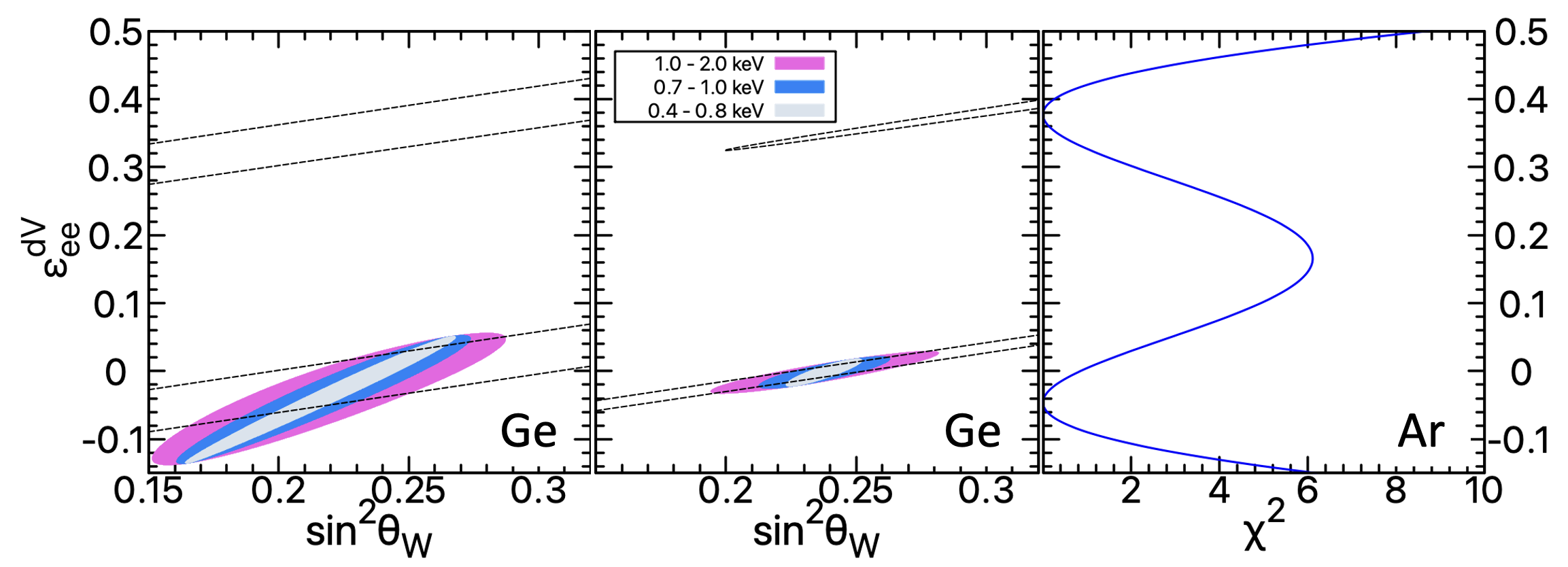}
\end{center}
\caption{ Allowed values at a 90\% C.L.  for $sin^{2}\theta_{W}$ vs $\varepsilon_{ee}^{dV}$ (left), and $\varepsilon_{ee}^{dV}$ vs $\varepsilon_{\tau e}^{dV}$ (right). In both cases the region between dashed lines represents the case where no correlations are considered. The filled curves indicate the correlated case for the three different nuclear recoil energy regions under study for reactor neutrinos.  We consider the two main systematic error contributions as $\sigma_{k}^{A} =$ 25\% and $\sigma_{k}^{B} =$ 10\% (left) and 5\% each (central). Background effects are considered by adding a 10\% contribution of the SM prediction to the statistical uncertainty.
  \label{fig:WMA-NSI}}
\end{figure}

Besides constraining parameters that are relevant for nuclear physics,
we can perform tests to the SM at low energies, as well as we can also
get information about new physics scenarios.  In the case of the SM,
the weak mixing angle appears in the cross section of CEvNS through
the relation:

\begin{equation}
g_{V}^{p} = \frac{1}{2} - 2\sin ^{2}\theta_{W}
\end{equation}

On the other hand, we consider the formalism of Non-Standard Interactions (NSI) as the parametrization of new physics.  Including these effects, the CEvNS cross section for an incoming electron (anti)neutrino now reads \cite{Barranco:2005yy}:

\begin{equation}
\begin{aligned}
\frac{\mathrm{d} \sigma }{\mathrm{d} T}\left ( E_{\nu },T \right ) \simeq & \frac{G_{F}^{2}M}{\pi }\left ( 1-\frac{MT}{2E_{\nu }^{2}} \right )\left \{ \left [ Z\left ( g_{V}^{p}+2\varepsilon _{ee}^{uV}+\varepsilon _{ee}^{dV} \right )F_{Z}^{V}(q^{2})+N\left ( g_{V}^{n}+\varepsilon _{ee}^{uV}+2\varepsilon _{ee}^{dV}  \right ) F_{N}^{V}(q^{2})\right ]^{2}\right.\\
& \left. +\sum_{\alpha = \mu, \tau}\left [ Z\left ( 2\varepsilon _{\alpha e}^{uV}+\varepsilon _{\alpha e}^{dV} \right )F_{Z}^{V}(q^{2})+N\left ( \varepsilon _{\alpha e}^{uV}+2\varepsilon _{\alpha e}^{dV} \right )F_{N}^{V}(q^{2}) \right ]^{2} \right \} 
\end{aligned}
\label{CrossNSI}
\end{equation}

Where $\varepsilon_{\alpha\beta}^{fV}$ are the constants that parametrize new physics. This expression contains non-universal ($\alpha = \beta$), as well as flavor changing ($\alpha\neq\beta $) parameters. In this way, we parametrize physics beyond the SM. The specific meaning of each of the couplings depends on the model under study. These include SM extensions like those considering two Higgs triplets, light mediators, non-unitarity of the mixing matrix, etc.  The presence of one or more of these parameters within the cross section, can induce a degeneracy in the allowed values of the weak mixing angle, as well as in the values of NSI's when considering two of them to be non-zero at a time. This degeneracy can be broken when using our proposed array of detectors. Here we exemplify with two different cases. To this end, we consider reactor neutrinos, since they are on an energy regime, where the form factor does not play a significant role, which allows us to study NSI parameters in a cleaner  manner. Again, we perform a $\chi^{2}$ analysis assuming the two different configurations of systematic error contributions.

Fig. (\ref{fig:WMA-NSI}) shows the analysis when we vary the weak
mixing angle and one non-universal NSI parameter
($\varepsilon_{ee}^{dV}$). The regions enclosed by dashed lines are
the allowed regions at a 90\% C.L. when no correlations between
detectors are present. Notice that here we have two different regions,
allowing values for $\sin^{2}\theta_{W}$ even smaller than 0.1 and
larger than 0.4, while the NSI can take relatively large values around
0.4.  In the same figure, we show the allowed regions when we consider
correlations between detectors, where we have separated our results
according to the three regions of the differential recoil energy
defined in section \ref{sec:N}.  The magenta region shows the results
for the high energy tail of the spectrum $1 < T < 2$ keV, the blue
region for $0.7 < T < 1$ keV, and the gray region for the low energy
threshold $0.4 < T < 0.8$ keV.  We notice that the degeneracy of the
allowed values for these parameters is broken for both of the
systematic error contribution scenarios. For comparison, we show the
$\chi^{2}$ analysis for the current Ar data when fixing the weak
mixing angle to the low-energy central value $\sin ^{2}\theta_{W} =
0.23865$ \cite{Miranda:2020tif}. We notice that the degeneracy in the
NSI parameter is also present in the Ar result. We conclude that our
proposed method can be used to break this degeneracy by using a single
measurement by taking advantage of the correlation between different
isotopes.

\begin{figure}[t] 
\begin{center}
\includegraphics[width=0.8\textwidth]{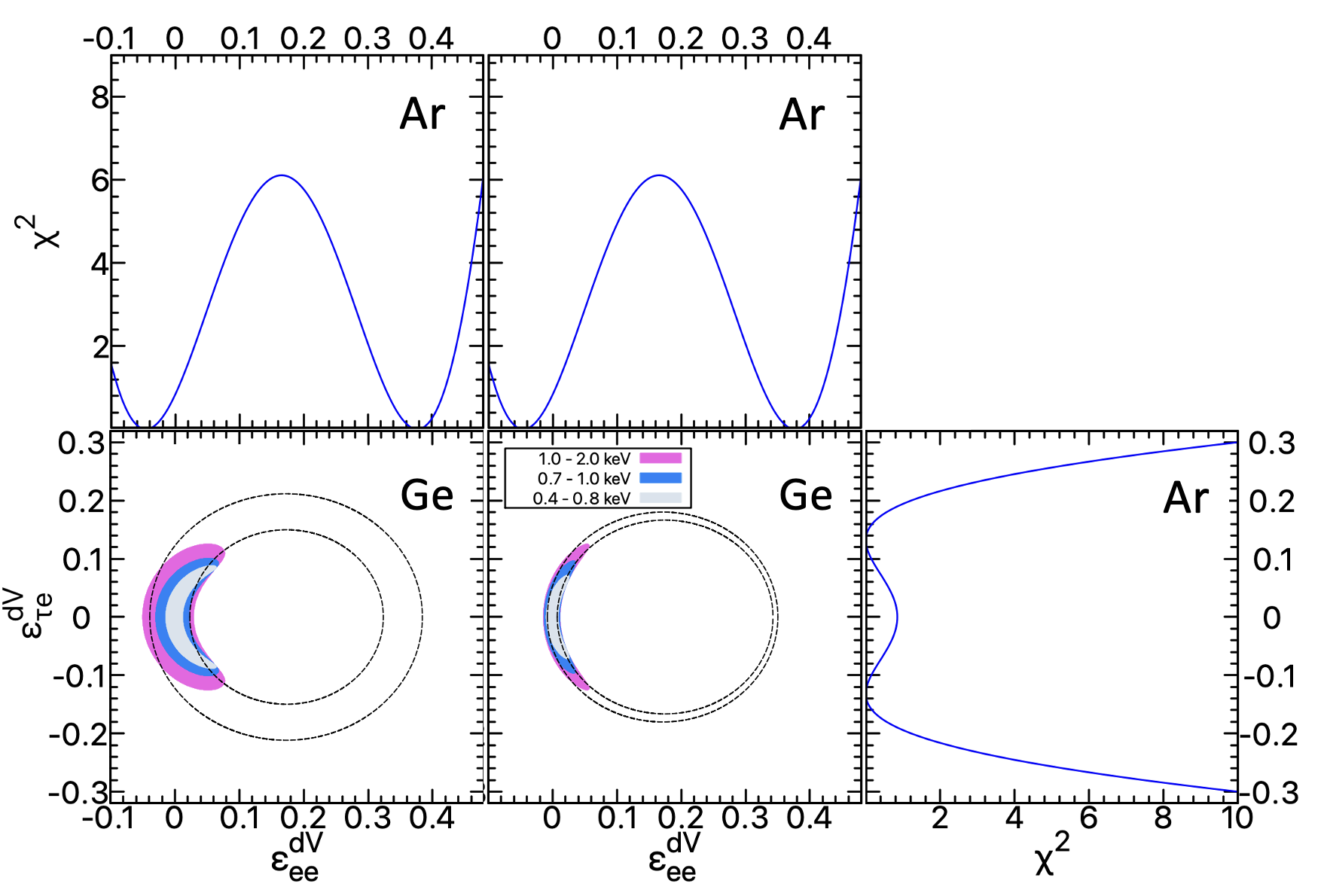}
\end{center}
\caption{ Allowed values at a 90\% C.L.  for $sin^{2}\theta_{W}$ vs $\varepsilon_{ee}^{dV}$ (left), and $\varepsilon_{ee}^{dV}$ vs $\varepsilon_{\tau e}^{dV}$ (right). In both cases the region between dashed lines represents the case where no correlations are considered. The filled curves indicate the correlated case for the three different nuclear recoil energy regions under study for reactor neutrinos.  We consider the two main systematic error contributions as $\sigma_{k}^{A} =$ 25\% and $\sigma_{k}^{B} =$ 10\% (left) and 5\% each (central). Background effects are considered by adding  a 10\% contribution of the SM prediction to the statistical uncertainty.
  \label{fig:NSI-NSI}}
\end{figure}

A similar analysis is show in Fig. (\ref{fig:NSI-NSI}), this time
considering the weak mixing angle fixed and varying two different NSI
parameters at a time, one non-universal, and one flavor changing. The
region between dashed lines shows the case when no correlations are
considered, while the colored regions show the results when
correlations are present under the same color code as defined in the
previous case. We notice that, again, the degeneracy is broken between
parameters. For comparison, we also show the current analysis for Ar
data when assuming only one of each of the parameters to be non-zero
at a time \cite{Miranda:2020tif}.  We notice that, again, the current
degeneracy on the non-universal parameter can be broken by using an
array of different isotopes.

\section{conclusions}
In summary we present a novel approach for precision measurements of
CEvNS. There are no major obstacles for constructing and operating
multi-isotope detector arrays based on current technologies. Detailed
sensitivity studies will require of the knowledge of the detector
response and background models associated with the specific
experimental conditions. Second order effects in the detectors
attributed to differences in the isotopic composition of its
constituents will have to be taken into consideration such as
departures from 100\% enrichment, differences in fiducial masses,
differences of quenching factors and cosmogenic and neutron-induced
backgrounds~\cite{Mayer:1970,Barabanov:2005cw}.

We have shown that measuring CEvNS from the same neutrino source at
the same time with several isotopically enriched detectors will
improve the accuracy of the measurement.  We have illustrated this
idea with a well motivated array based on three different germanium
isotopes and applied to the concrete example of confirming the $N^2$
rule.  We have also illustrated how, in the case of a $\pi$-DAR
neutrino flux, the measurement of the mean neutron radius can be
improved. For the reactor antineutrino case, we have shown how the
proposed array can provide a robust test for new physics The proposal
discussed here will help to achieve the necessary accuracy to untangle
different contributions from nuclear and particle physics allowing for
a reliable constraint on physics beyond the SM that would be free from
neutron charge distribution
uncertainties~\cite{AristizabalSierra:2019zmy} and can resolve the
characteristic degeneracies appearing in these scenarios.  The same
technique can be applied to other technologies that allow lower energy
thresholds as in the case of bolometers.  After the first detection of
CEvNS at the SNS, the next generation of experiments will aim for
precision measurements. Despite several years and a worldwide effort,
detection of CEvNS at a reactor has not been observed.  Neutrino
fluxes and quenching factors remain as considerable sources of
uncertainty. We have shown that with the current level of
uncertainties, this proposed array would allow a clear signature of
(anti)neutrino detection, keeping the different systematics under
control.  Our combinations of systematic error contributions show how
future improvements on the systematics will improve the accuracy of
the measurements for different observables.

This work was performed under the auspices of the U.S. Department of
Energy by Oak Ridge National Laboratory under Contract No. DE-AC05-
00OR22725. This work has been supported by CONACyT under grant
A1-S-23238. The authors appreciate discussions with J. Collar, Chicago, and B. Littlejohn, Illinois Tech.

\end{document}